\documentclass{article}

\usepackage{arxiv}

\usepackage[utf8]{inputenc} 
\usepackage[T1]{fontenc}    
\usepackage{hyperref}       
\usepackage{url}            
\usepackage{booktabs}       
\usepackage{amsfonts}       
\usepackage{nicefrac}       
\usepackage{microtype}      
\usepackage{lipsum}

\usepackage{amsmath}
\usepackage{amssymb}
\usepackage{dcolumn}
\usepackage[utf8]{inputenc}
\title{Explaining Agent-Based Financial Market Simulation}

\author{
  David Byrd\\
  School of Interactive Computing\\
  Georgia Institute of Technology\\
  Atlanta, GA 30308\\
  \texttt{db@gatech.edu}\\
}

\begin{document}
\maketitle

\begin{abstract}
This paper is intended to explain, in simple terms, some of the mechanisms and agents common to multiagent financial market simulations.  We first discuss the necessity to include an exogenous price time series ("the fundamental value") for each asset and three methods for generating that series.  We then illustrate one process by which a Bayesian agent may receive limited observations of the fundamental series and estimate its current and future values.  Finally, we present two such agents widely examined in the literature, the Zero Intelligence agent and the Heuristic Belief Learning agent, which implement different approaches to order placement.
\end{abstract}

\section*{The True Fundamental Value Times Series}

A common feature of agent-based financial simulation is the presence of an exogenous (\emph{defined:} ``derived from external factors'') price time series, representing some global consensus value for an asset at any time, arising from the accumulation of all available news and information which could influence the valuation of the asset.  A simple intuition for the need of such an extrinsic value can be obtained by its omission.  Imagine a simulated stock entering its opening auction period, such that there is not yet a limit order book to examine.  In the absence of an extrinsic value of some kind, what would cause market agents to place bids or offers with any particular limit price?  We could imagine using the prior day's closing price as an anchor, but then what about the first day?  There must be \emph{some} concept of ``what this company is really worth'' for at least \emph{some} of the market participants, or there is no reason for the stock to trade in the \$5 range versus the \$5,000 range.  In real-world terms, we can imagine consultation of this extrinsic series to be the distinguishing factor between value investors versus purely technical traders.

This ``fundamental value series'' requires an omniscient perspective that is not available to any market participant.  Thus at \emph{best}, market participants may obtain a noisy observation of this value at some particular point in time.  This paper focuses on the set of mean-reverting fundamental value series with characteristics as described in Section 3 of Chakraborty and Kearns \cite{chakraborty2011market}.

\subsection*{Properties of the Fundamental}

Intuitively, think of the fundamental as a consensus valuation of all people everywhere that somehow reflects the ``real'' value of a company.  This value changes over time in response to events, and of course no one can ever know precisely what the global consensus value is.

\begin{itemize}
    \item There is exactly one fundamental value time series for each equity in the simulation.
    \item It is a property of the equity and does not vary per agent.
    \item It is \emph{predetermined}; agent actions do \emph{not} affect it.
\end{itemize}

Most any type of numeric time series data can serve as a fundamental value series for an equity.  For example, one could simply use historical market data.  We consider three possibilities: a discrete mean reverting series, a continuous mean reverting series based on the Ornstein-Uhlenbeck process, and a modification of that OU process with ``megashock events''.

\subsection*{Discrete Mean Reverting Fundamental}

We first consider a discrete mean reverting fundamental time series as presented in Section 4 of Wah et al.  \cite{wah2017welfare}

In this case, the fundamental value series for an equity is defined at all times $t\in[0,T]$ by:
$$r_t=\mathbf{max}\{0,\kappa\bar{r}+(1-\kappa)r_{t-1}+u_t\};r_0=\bar{r}$$

The mean fundamental value for an equity ($\bar{r}$) is chosen as part of experimental configuration.  The fundamental value series $r$ begins at this mean ($r_0=\bar{r}$) and is never allowed to become negative.

At each time step $t \in [0,T]$, the series makes a single-step reversion to the mean using parameter $\kappa\in[0,1]$, the mean reversion rate.  This functions similarly to the learning rate in many machine learning algorithms: $\alpha\times\mathtt{new\_value}+(1-\alpha)\times\mathtt{old\_value}$.  Here, \texttt{new\_value} is just $\bar{r}$ (to produce mean reversion).  Thus the series is always drifting back toward the mean at a predictable rate, faster for high $\kappa$ and slower for low $\kappa$.  If $\kappa = 0$, of course, the time series does not revert to the mean.

So far this looks \emph{very} unlike a stock price.  When the price deviates from the mean, it slowly and predictably returns to the mean, then stays there forever.  In fact, nothing thus far would make it leave the mean in the first place.

Enter $u_t\sim\mathcal{N}(0,\sigma^2_s)$, the ``shock variance''.  This is a random number drawn at each time step $t$ to perturb the otherwise completely predictable fundamental value series.  It erratically pushes the series around (including away from the mean) to introduce noise into the price series and create a random walk.

\textbf{Important:} This $u_t$ ``noise'' is part of producing the single, true fundamental value series for an equity.  This is \emph{not} the observation noise experienced by agents.

Agents are sometimes required to infer the fixed parameters of the mean reverting fundamental, but for simplicity it is often assumed that agents have become ``tuned'' to their environment.  In either case, agents \emph{do} know the above equations, but do not know the true fundamental value after time $t=0$, because of the random draws for $u_t$.  In the ``tuned'' case, agents additionally know that $u_t\sim\mathcal{N}(0,\sigma^2_s)$, and know the correct values for $\bar{r}$, $\kappa$, and $\sigma^2_s$.

\subsection*{Ornstein-Uhlenbeck Process Fundamental}

We next consider a continuous mean reverting time series as the fundamental, based on the Ornstein-Uhlenbeck (OU) Process as described in Section 3.1 of Chakraborty and Kearns.  \cite{chakraborty2011market}

A potential weakness in the discrete mean reverting series is that it must be computed at every discrete time step, in order, for $t \in [0,T]$.  Time steps at which no agent activity occurs must still be evaluated to obtain future values.  If there are many time steps, and agents tend to arrive at the market infrequently, this can become a significant computational burden for the simulation.

The OU Process, with values represented by $Q$, has the advantage that $Q_t$ can be computed for any future time $t$, given value $Q_0$.  Because OU has the Markov property (\emph{i.e.} $Q_{t+1}$ is conditionally independent of $Q_{t-1}$ given $Q_t$) and $Q_t$ is normally distributed, we can determine the expectation and variance of $Q_t$ and obtain the value by sampling, without computing any of the values ``in between''.  In the case of sparse agent arrivals, this can save considerable computation effort for the simulation.

The OU process requires similar parameters and values to the discrete mean reverting process.  The equivalent discrete series parameter is given in parenthesis after each OU compoment.  Specifically, the OU process requires a mean fundamental $\mu$ (equiv: $\bar{r}$), a prior value $Q_0$ (equiv: $r_{t-1}$), a mean reversion rate $\gamma$ (equiv: $\kappa$), and a volatility value $\sigma$ (equiv: $\sigma^2_s$).

The value $Q_t$ of the OU process at time $t$ can then be sampled from a normal distribution with mean: $\mu + (Q_0 - \mu)e^{-\gamma t}$, and variance: $\frac{\sigma^2}{2\gamma}(1-e^{-2\gamma t})$.

Like the discrete mean reverting process, the OU process can only move away from the mean through its volatility value.  If this value is kept small to represent high-frequency ``noise'' in the process, both processes will appear as essentially single-scale accumulating white noise around the fundamental mean, and will not look very much like a plot of, for example, a day of real transactions (or quotes) for some equity traded on a US exchange.

\subsection*{Megashock OU Fundamental}

To obtain the simulation computation time improvements of the OU Process Fundamental, while producing a series that looks more like a real market equity price time series, we introduce the concept of ``megashock events''.  Megashock events are intended to represent extrinsic news of a substantial nature that occur relatively infrequently, but have the potential to significantly alter the consensus valuation of a stock.

In technical terms, megashock events are layered on top of the preliminary OU fundamental, arriving via a Poisson process, and are drawn from a bimodal distribution with mean zero.  Each sub-distribution is a Gaussian with mean significantly different from zero, and a high variance relative to the OU process base variance.  The sub-distributions are centered in positive and negative territory with mean equally far from zero, producing the desired overall mean zero for the bimodal distribution.

\section*{Estimating the Final Fundamental Value}

In the simulation described in Wang and Wellman \cite{wang2017spoofing}, agents attempt to maximize the value of their portfolio at the end of the simulation relative to the final fundamental value.  This is what the closing price ``should'' be.  It is \emph{not} the same as marking to market.  Regardless, predicting this final fundamental value is an important part of making trading decisions.

The agents arrive according to a Poisson distribution and a discrete mean reverting process is used for the fundamental.  Assume some agent wakes at time $t$ and receives new (noisy) fundamental observation $o_t$.  This observation is unique to each agent with $o_t=r_t+n_t$, where $r_t$ is the true fundamental value and $n_t$ is random observation noise drawn from $\mathcal{N}(0,\sigma^2_n)$.  The agent \emph{does} know $\sigma^2_n$.

The agent maintains an estimate $\tilde{r}_t$ of the current fundamental value $r_t$, and an estimated variance $\tilde{\sigma}^2_t$ \emph{of this estimate}.  \textbf{Important:} Again, $\tilde{\sigma}^2_t$ is the agent estimating the variance \emph{of its $\tilde{r}_t$ estimate}.  Thus, $\tilde{\sigma}^2_t$ is like an internal error metric (or loosely an ``inverse confidence''), not an estimate of any market or simulation parameter.

Let $t'$ be the last time the agent woke.  At that time, it had some estimates ($\tilde{r}_{t'}$ and $\tilde{\sigma}_{t'}$) of $r_{t'}$ and its own estimation error.  First it must mentally advance time from $t'$ to $t$.  With no observations while asleep, it can only apply mean reversion at each time step:
$$\tilde{r}_{t'} \leftarrow (1-(1-\kappa)^\delta)\bar{r} + (1-\kappa)^\delta \tilde{r}_{t'}$$
In the above update, $\delta=t-t'$, the number of time steps since the agent last woke.
Note that with $\delta=1$, it is exactly the mean reversion equation used to compute the true fundamental value series $r_t$, except: the agent does not know $r_{t-1}$ and so uses its previous estimate $\tilde{r}_{t'}$ instead, and the agent does not know the random perturbations $u_t$ and so uses the known mean value of zero instead.

With $\delta>1$, this is equivalent to applying mean reversion $\delta$ times.  The first time the agent wakes, its previous wake time is effectively $t'=0$.  Thus it should assume $\tilde{r}_{t'}=\bar{r}$, as it knows the value of $\bar{r}$ and that $r_0=\bar{r}$.

To complete the advancement of time from $t'$ to $t$, the agent must also update its internal error estimate $\tilde{\sigma}_{t'}$, which is initially zero.  It does so as follows:
$$\tilde{\sigma}^2_{t'} \leftarrow (1-\kappa)^{2\delta} \tilde{\sigma}^2_{t'} + \frac{1-(1-\kappa)^{2\delta}}{1-(1-\kappa)^2} \sigma^2_s$$
This update is a weighted mixture of the previous internal error estimate $\tilde{\sigma}^2_{t'}$ and the shock variance $\sigma^2_s$.  The weight term for shock variance $\sigma^2_s$ starts at $1$ when $\delta=1$ and increases with $\delta$.  The weight term for previous internal error estimate $\tilde{\sigma}^2_{t'}$ starts at $(1-\kappa)^2$ when $\delta=1$ and decreases as $\delta$ increases.  Thus when computing the new error estimate, the more time that has passed since the agent's last observation, the more weight is placed on shock variance $\sigma^2_s$ and the less weight is placed on the prior error estimate $\tilde{\sigma}^2_{t'}$.

Having now advanced time from $t'$ to $t$, the agent must apply its new observation $o_t$.  This process is not unlike a Dynamic Bayes Network, in which the agent alternates stepping forward in time (transition model) and applying new evidence (sensor model).  The current estimate of the fundamental value is a weighted combination of the prior estimate with the new observation:
$$\tilde{r}_t=\frac{\sigma^2_n}{\sigma^2_n+\tilde{\sigma}^2_{t'}}\tilde{r}_{t'} + \frac{\tilde{\sigma}^2_{t'}}{\sigma^2_n+\tilde{\sigma}^2_{t'}}o_t$$
The weights are simply the observation noise $\sigma^2_n$ and the agent's internal estimate of its error $\tilde{\sigma}^2_{t'}$, normalized to sum to one.  The observation noise weights the agent's prior estimate of fundamental value $\tilde{r}_{t'}$ and the internal error estimate weights the new observation $o_t$.

\textbf{Important:} The above weights are reversed from what might be initially expected.  This is appropriate because a higher variance term is resulting in a \emph{higher} weight, which is the opposite of what we would want, so we switch the weights.  The higher our estimation of our internal error, the more weight we put on the new observation.  The higher the observation noise, the more weight we put on our previous estimates.

The agent must also finish updating its internal error estimate for use the next time it wakes:
$$\tilde{\sigma}^2_t=\frac{\sigma^2_n\tilde{\sigma}^2_{t'}}{\sigma^2_n+\tilde{\sigma}^2_{t'}}$$
Similarly to applying the new observation to $\tilde{r}_t$, here we apply one ``step'' of the observation noise $\sigma^2_n$ to our internal error estimate $\tilde{\sigma}^2_t$.  Making an observation of course tends to reduce our internal error estimate (or increase our ``certainty'') but with higher observation noise, this effect is diminished.

The agent now has updated estimates $\tilde{r}_t$ and $\tilde{\sigma}^2_t$ for the current fundamental value and its internal error estimation.  At last, it can compute $\hat{r}_t$, the final fundamental value $r_T$ as estimated at current time $t$.  Having no future observations (past time $t$) it can only advance time as before:
$$\hat{r}_{t} \leftarrow (1-(1-\kappa)^{T-t})\bar{r} + (1-\kappa)^{T-t} \tilde{r}_{t}$$
This matches the previous ``time advancement'' process with $\delta=T-t$, the number of time steps remaining in the simulation.  Again the agent must assume the random perturbations $u_t$ take on their mean value of zero.

Now the agent can use its estimate of the final fundamental value (or what the closing price \emph{should} be) to make its trading decisions.

Note that the above estimation method is not substantially different when the fundamental follows the OU process, because the two processes map very closely.

\section*{Agent Preferences}

It is important that each agent have individual preferences for holding certain quantities of stock, as we would want each agent to behave differently even without forcing arbitrary randomness into the decision process.  In the real world, independent of any ``consensus value'' for INTC, some people will just \emph{really} want to own Intel.  Here we  describe the method as put forth in Wah and Wellman.  \cite{wah2017welfare}

These preferences are codified as a vector of \emph{incremental private values} placed on the acquisition or release of one additional unit of stock, given the agent's current holdings.  If $q_{max}$ is the holding limit, then the preferences for trading agent $i$ are the elements $\theta^q_i$ in:
$$\Theta_i=(\theta^{-q_{max}+1}_i,\dots,\theta^0_i,\theta^1_i,\dots,\theta^{q_{max}}_i)$$
where $q$ is the quantity of stock currently held.  $\Theta_i$ is drawn randomly from $\mathcal{N}(0,\sigma^2_{PV})$, where $\sigma^2_{PV}$ is a selected experimental parameter.  The values are sorted in descending order, ensuring that each additional unit of stock acquired is valued less than the one before it.

For example, consider a private value vector for some particular agent:
$$\Theta^q=(0.5,0.3,0.2,0.1,-0.2,-0.4)$$
for $q\in[-2,3]$, and assume this agent's $\hat{r}_t$ estimate of the final fundamental value is $100$.  Since each agent's total valuation for its $q\textsuperscript{th}$ unit of stock is $\hat{r}_t + \theta^q$, we can compute the following total valuations:
$$V^q=(100.5,100.3,100.2,100.1,99.8,99.6)$$

If the agent currently holds zero units of stock, it would pay (\emph{see caveat in next paragraph}) \$100.10 to go long one unit, or demand \$100.20 to go short one unit.  If the agent is currently short two units, it would pay \$100.30 to buy back a unit, or demand \$100.50 to sell another unit.  If the agent is long two units, it would pay \$99.60 to buy a third unit, or require \$99.80 to sell a unit.

It is this total valuation of a unit (incremental private preference plus estimated final fundamental value) that governs each trader's limit prices on its orders.  However an agent should not bid/offer \emph{exactly} the limit prices listed above, as that would result in zero surplus (gain) versus its valuation.  Instead it will \emph{shade} its bids lower or offers higher to ensure a positive surplus if the order is filled.

\section*{Zero Intelligence Agent}

The Zero Intelligence (ZI) agent has been a part of the financial literature from at least 1993, when Gode and Sunder studied the allocative efficiency of constrained or unconstrained ZI agents with that of a population of human traders, finding that efficiency arose as a consequence of market structure rather than human intelligence or motivation.  \cite{gode1993allocative}  Today, the term Zero Intelligence (or Zero Intelligence Plus) is extended to more complex agents, so long as they place orders priced substantially at random, at times selected substantially at random, and do not possess significant memory, intelligence, or visibility into the order stream.

Here we again describe the particular flavor of ZI agent used in Wah and Wellman.  \cite{wah2017welfare}  We note that these agents are not \emph{strictly} zero intelligence, because they (1) do observe the current best bid or ask as part of a final rejection process when placing an order and (2) the same agent arrives at the market multiple times, versus only once in ``classic'' ZI.

Upon waking, a ZI trader cancels any outstanding order and places a new, single-unit limit order with equal probability to buy or sell.

The total valuation used by a ZI trader for stock unit $q$ is always as explained in the previous section: $\hat{r}_t+\theta^q$.  If the ZI agent currently holds $q$ units, then it could contemplate selling unit $q$ with valuation $\hat{r}_t+\theta^q$ or buying unit $q+1$ with valuation $\hat{r}_t+\theta^{q+1}_i$.

To ensure a surplus (gain) from the trade, the agent must \emph{shade} its limit prices away from this valuation, as pricing at the valuation would produce zero net gain even if the order is executed.  Bids are shaded downward (bid less than your valuation) and asks are shaded upward (ask more than your valuation).

Call the trader's total valuation for the share in question $v$.  The range $[R_{\mathrm{min}},R_{\mathrm{max}}]$ represents the minimum and maximum surplus (gain) the agent will demand on a filled order.  The agent draws a random number $R$, called the \emph{requested surplus}, from that range.  $R$ is the difference between the agent's total valuation of the stock unit and the ``shaded'' limit price it actually offers.  The agent will gain $R$ surplus if the order trades, or nothing if it does not.  Thus the limit price is $v-R$ if buying, and $v+R$ if selling.

This particular (Wah and Wellman \cite{wah2017welfare}) agent also has a specified experimental parameter $\eta\in[0,1]$ called the ``strategic threshold''.  If an agent can immediately secure an executed surplus (gain) of $\eta R$ by taking the current best bid or offer, it will do that instead of placing the order previously described.

\textbf{Shortcoming:} Note that the ZI agent selected a ``requested surplus'' $R$ at random with some fixed parameter range.  This may well result in its order not being executed at all (too far from the spread).  \emph{This} is the problem solved by the Heuristic Belief Learner (HBL) agent, which improves overall performance by estimating the likelihood that any particular limit price will successfully trade, therefore allowing it to bid less or ask more \emph{without} too badly damaging the odds that the order will be filled.

\section*{Example of ZI Limit Price Determination}

Here we extend the private and total valuations example from the \emph{Agent Preferences} section, using private value vector:
$$\Theta^q=(0.5,0.3,0.2,0.1,-0.2,-0.4)$$
and total value vector:
$$V^q=(100.5,100.3,100.2,100.1,99.8,99.6)$$
for $q\in[-2,3]$, and assuming $\hat{r}_t=100$ as before.

An agent holds (long) $q=1$ unit of stock.  It has been randomly selected to buy on this market arrival.  The agent estimates the final fundamental at 100.  Its private valuation for stock unit $q+1=2$ is $-0.20$, thus its total valuation for stock unit 2, buying one unit while already holding one unit, is 99.8.

Assume $R_{\mathrm{min}}=0.1$ and $R_{\mathrm{max}}=0.5$.  This means the agent will place orders with limit prices chosen to achieve at least 10 cents gain (for one unit) but no more than 50 cents gain relative to its total valuation of the stock unit.  Assume the agent randomly selects $R$ to be 0.25 this time.  Since the agent is buying, its BID limit price will be $99.8-0.25=99.55$, which will net a gain of 0.25 if the order is filled.

The strategic threshold $\eta$ \emph{could} cause the agent not to place the above order.  Assume $\eta=0.5$, meaning the agent will accept one-half of its desired gain if the order can be executed immediately.  If the agent can get at least $\eta R=0.5(0.25)=0.125$ by accepting the current best offer, it will do that.  In this example, if the best offer is at most 99.67, the agent will place a bid to trade with that offer.  Otherwise, it will place the bid with limit price 99.55 as planned.

\section*{Heuristic Belief Learning Agent}

GD \cite{gjerstad1998price}, named for its authors Gjerstad and Dickhaut, is at the root of a family tree of strategies that maintain a belief function representing the probability that a particular auction bid will be accepted based on its price.  Heuristic Belief Learning (HBL) is the name used for a generalized form of GD presented later by Gjerstad.  \cite{gjerstad2007competitive}  Here we present HBL as described in Wang and Wellman.  \cite{wang2017spoofing}

These HBL agents work exactly like the previously described ZI agents (having private preferences, arriving according to a Poisson distribution, estimating the final fundamental value, etc) except as noted herein.

Using its belief function (likelihood of bid success conditioned on price), an HBL agent places the bid that maximizes expected surplus, which accounts for both the surplus \emph{if} the bid is accepted and the likelihood the bid will \emph{be} accepted.  This belief function is constructed from an accurately observed history of all bids (accepted or rejected) leading to the last $L$ (memory length) transactions.  In the context of a more complex market with an order book that allows delayed execution and order cancellation, we can instead estimate the probability that a limit order will be successfully transacted within some maximum time period based on its limit price.

\textbf{Important:} This belief function has \emph{nothing} to do with price prediction, \emph{nothing} to do with the ``true value'' of an item, and \emph{nothing} to do with the profitability of a transaction.  It simply asks: ``If I were to place a bid at this price, what are the odds I would receive the item I bid on?''

GD and HBL are considered heuristic agents because, as we will see, their estimation of bid success probability uses an arbitrary (but intuitive) formulation.

\section*{Intuitive Formulation of HBL Success Likelihood Estimation}

At any given time, the HBL agent can use its memory of observed limit orders and their results to estimate the probability of a successful transaction given any observed price.

If the HBL wished to buy, we could construct a simple ratio of successful to unsuccessful bids at exactly the proposed limit price $p$, but this ratio is unbounded and we want a probability.  Normalizing the denominator restricts the range to $[0,1]$ and permits our desired interpretation:
$$Pr(p)=\frac{S_p}{S_p+U_p}$$
where $S_p$ is the number of successful bid orders at price $p$, and $U_p$ is the number of unsuccessful bid orders at price $p$.  The agent thus asks: ``Of all bids at price $p$, what proportion were successful?''

This approach has an obvious limitation: data sparsity.  If there have been ten unsuccessful bids at $p=5$, ten successful bids at $p=7$, and ten successful bids at $p=9$, the value $Pr(8)=\frac{0}{0+0}$ is undefined.  If we extend $S$ and $U$ to include bids equal or less than $p$:
$$Pr(p)=\frac{S_{\leq p}}{S_{\leq p}+U_{\leq p}}$$
the agent now asks: ``Of all bids at or below price $p$, what proportion were successful?''.  Now $Pr(8)=\frac{10}{10+10}=0.5$, which is at least \emph{something}.

Including unsuccessful bids at prices \emph{lower} than our proposed price is unhelpful, though.  If we are offering 8, why should failed bids at 5 affect our estimation of success?  $Pr(8)=0.5$ seems unrealistic; \emph{all} bids at $p=7$ were \emph{accepted}!

Instead, we should evaluate successful bids at or below $p$, but unsuccessful bids at or \emph{above} $p$.  The agent now considers two different factors: ``How many bids succeeded offering no more than this?'' and ``How many bids failed \emph{despite} offering at least this much?''
$$Pr(p)=\frac{S_{\leq p}}{S_{\leq p}+U_{\geq p}}$$
We now appropriately disregard the failed bids at 5, because we are offering more than that, and estimate $Pr(8)=\frac{10}{10+0}=1$, because we have never observed the failure of a bid with a price of at least 8.

There is still one problem.  Orders in our type of auction can be cancelled.  Consider the ten unsuccessful bids at $p=5$.  Were they unsuccessful because there has never been an ask at $p<=5$?  Or were they merely unlucky, and there have been offers that low, but not at the same time as those bids?  In the second case, our estimation of $Pr(5)$ should be greater than zero, because although no bid succeeded, one clearly \emph{could} have.

Our agent also has limited memory.  What if it should happen that within this memory, there were no bids at or below $p$ (successful or not), but there were ask orders below $p$?  Our heuristic might end up zero or undefined, when again, the chance of fulfillment is near certain due to the volume of unmatched ask orders.

Generally speaking then, the volume of asks in our considered price range ($<=p$) should be a factor.  In addition to considering bids and their success rate, the more shares that have been \emph{available} at or below $p$, the more likely we are to succeed with a bid at $p$.  The fewer shares that have been available, the less likely we are to succeed.

Thus we arrive at the final formulation for the HBL agent's heuristic estimation of success, with this rough interpretation: ``Seeing ask order volume, or successful bid orders, at prices no more than $p$ increases the likelihood of success.  Seeing unsuccessful bid orders at prices of at least $p$ decreases the likelihood of success.''  The ratio is formalized and normalized as:
$$Pr(p)=\frac{A_{\le p}+S_{\leq p}}{A_{\le p}+S_{\leq p}+U_{\geq p}}$$
where $A_{\leq p}$ is the total number of ask orders at price $p$ or lower, $S_{\leq p}$ is the number of successful bid orders at price $p$ or lower, and $U_{\geq p}$ is the number of unsuccessful bid orders at price $p$ or higher.  The heuristic calculation is exactly mirrored when the agent considers selling a unit.

The agent can now estimate the probability of success for any candidate limit order price $p$.  If it is desired to estimate probability for all possible prices without bound, cubic spline interpolation can be employed as suggested by Wang and Wellman. \cite{wang2017spoofing}

\newpage

\section*{HBL Considerations and Strategy}

In a market with an order book, orders are almost never rejected.  They simply enter the order book to await a later match.  The HBL, though, depends on the idea of rejected or unsuccessful orders.  A simple approach to this problem would be to place a time limit on each order.  If the order is not matched within this time limit, it is considered unsuccessful.

The approach taken by Wang and Wellman is a little more complex. \cite{wang2017spoofing} They define a grace period relative to the frequency of trader arrivals (the Poisson distribution mentioned earlier).  An order is considered rejected if it existed in the order book for a length of time greater than the grace period.  All orders not accepted immediately are considered ``partially rejected'' and assigned a fractional weight depending on the length of time they spent in the order book.

The HBL agent selects the limit price that maximizes \emph{expected surplus}, where surplus is simply the difference between the limit price $p$ and the agent's total valuation of a share (see ``Agent Preferences''), and expected surplus is the same, weighted by the probability of successful transaction computed in the previous section.

For example, an HBL agent contemplating buying unit $q+1$ would consider total valuation $\hat{r}_t+\theta^{q+1}_i$, limit price $p$, and estimation of successful transaction $Pr(p)$, and select its optimal bid $p^*$ as:
$$p^*=\mathrm{argmax}_p(\hat{r}_t+\theta^{q+1}_i-p)Pr(p)$$

Anytime the HBL agent does not have enough information to enact the above strategy, it temporarily behaves as a ZI agent instead.

\textbf{Limitation:} HBL assumes the agent can observe a complete order stream from all agents, including order execution (even if delayed), cancellation, and the length of time each order was ``alive'' in the order book.  This may not be practical in many environments.

\section*{HBL Limit Price Example}

Here we reuse the private and total valuations example from the \emph{Example of ZI Limit Price Determination} section, using private value vector:
$$\Theta^q=(0.5,0.3,0.2,0.1,-0.2,-0.4)$$
and total value vector:
$$V^q=(100.5,100.3,100.2,100.1,99.8,99.6)$$
for $q\in[-2,3]$, and assuming $\hat{r}_t=100$ as before.

An agent is short one unit of stock ($q=-1$).  It has been randomly selected to buy on this market arrival.  The agent estimates the final fundamental at 100.  Its private valuation for stock unit $q+1=0$ is $0.20$, thus its total valuation for stock unit 0, buying one unit while being short one unit, is 100.2.

Assume the HBL agent has access to the following order memory of length $L=4$, with superscripts denoting pairs of matched orders, and that no orders were cancelled:
\begin{center}
\begin{tabular}{c|D{.}{.}{-1}|c|D{.}{.}{-1}|c|D{.}{.}{-1}|c|D{.}{.}{-1}}
\multicolumn{2}{c|}{Transaction 1} & \multicolumn{2}{c|}{Transaction 2} & \multicolumn{2}{c}{Transaction 3} & \multicolumn{2}{c}{Transaction 4} \\
\hline
ASK & 100.0^1 & ASK & 100.2^2 & BID & 100.1 & BID & 100.2 \\
BID & 99.8 & BID & 100.0 & ASK & 99.9^3 & ASK & 100.4 \\
ASK & 100.3^4 & ASK & 100.3 & & & BID & 100.4^4 \\
BID & 99.6 & BID & 100.1^3 & & \\
BID & 100.0^1 & BID & 100.2^2 & & \\
\end{tabular}
\end{center}

The agent can compute the estimated probability of transaction for a proposed limit price $p$ as described in the previous section:
\begin{equation*}
  \begin{aligned}
  Pr(100.5)&=\frac{6+4}{6+4+0}=1 \\
  Pr(100.4)&=\frac{6+4}{6+4+0}=1 \\
  Pr(100.3)&=\frac{5+3}{5+3+0}=1 \\
  Pr(100.2)&=\frac{3+3}{3+3+1}=0.86 \\
  Pr(100.1)&=\frac{2+2}{2+2+2}=0.67
  \end{aligned}
  \hspace{1cm}
  \begin{aligned}
  Pr(100.0)&=\frac{2+1}{2+1+3}=0.50 \\
  Pr(99.9)&=\frac{1+0}{1+0+3}=0.25 \\
  Pr(99.8)&=\frac{0+0}{0+0+4}=0 \\
  Pr(99.7)&=\frac{0+0}{0+0+4}=0 \\
  Pr(99.6)&=\frac{0+0}{0+0+5}=0
  \end{aligned}
\end{equation*}

Now the agent can use its total valuation, the proposed limit price, and the estimated probability of transaction to compute expected surplus $\mathbb{E}(s)$:
\begin{equation*}
  \begin{aligned}
  \mathbb{E}(s|p=100.5)&=(100.2-100.5)(1)= -0.3 \\
  \mathbb{E}(s|p=100.4)&=(100.2-100.4)(1)= -0.2 \\
  \mathbb{E}(s|p=100.3)&=(100.2-100.3)(1)= -0.1 \\
  \mathbb{E}(s|p=100.2)&=(100.2-100.2)(0.86)= 0 \\
  \mathbb{E}(s|p=100.1)&=(100.2-100.1)(0.67)= 0.067 \\
  \mathbb{E}(s|p=100.0)&=(100.2-100.0)(0.5)= 0.1 \\
  \mathbb{E}(s|p=99.9)&=(100.2-99.9)(0.25)= 0.075 \\
  \mathbb{E}(s|p=99.8)&=(100.2-99.8)(0)= 0 \\
  \mathbb{E}(s|p=99.7)&=(100.2-99.7)(0)= 0 \\
  \mathbb{E}(s|p=99.6)&=(100.2-99.6)(0)= 0
  \end{aligned}
\end{equation*}

Intuitively, the agent can see that any price above \$100.2 would almost certainly be accepted, but the agent would lose money relative to its valuation of the stock unit.  The agent would love to achieve the surpluses associated with prices less than \$99.9, but it can guess that those bids would never transact.  Only the prices $p\in [99.9,100.1]$ have a nonzero chance of transaction \emph{and} a positive surplus if transacted.

In this example, the agent will place a limit buy order at limit price $p=100.0$, the price that produces the maximum expected surplus of \$0.10.

\section*{The ABIDES Simulator}

Our simulation platform, ABIDES (Agent-Based Interactive Discrete Event Simulation) provides the fundamental value series and agents discussed herein ``out of the box'', including the novel ``Megashock OU Fundamental''.  \cite{byrd2019abides}  It also provides the ability to use \emph{real} market data as the fundamental value series.  \cite{balch2019evaluate}  The ABIDES platform is available under a BSD-style license at \url{https://github.com/abides-sim/abides}.  The distribution includes relevant example configurations which are further explained in the project's wiki at \url{https://github.com/abides-sim/abides/wiki}.

The ABIDES simulator uses the OU process (plus megashock events) to provide a ``sparse discrete'' fundamental process that provides extremely fine time resolution (nanoseconds) while still permitting quick simulation of reasonable time scales (days) due to its ability to completely skip computation of time periods during which nothing happens.  The Megashock OU fundamental additionally produces intraday time series that more closely resemble typical real-world intraday stock charts.

The ABIDES implementation of Zero Intelligence agents reproduces the strategy described herein, including the strategic threshold parameter.  The implementation of Heuristic Belief Learning agents treats an order as successful if any part of it is transacted within the observed order stream, and unsuccessful otherwise.

\section*{Acknowledgements}
This material is based on research supported in part by the National Science Foundation under Grant no. 1741026, and by a JPMorgan AI Research Fellowship.

\bibliographystyle{unsrt}
\bibliography{references.bib}

\end{document}